\documentclass[journal,12pt,onecolumn]{IEEEtran}

\usepackage{amssymb}
\usepackage{epsfig}
\usepackage{algorithm}

\usepackage{fullpage}
\usepackage{cite}
\usepackage[cmex10]{amsmath}
\usepackage{mdwmath}
\usepackage{mdwtab}
\usepackage{caption}
\usepackage{verbatim}
\usepackage[tight,footnotesize]{subfigure}

\newcommand{\qed}{\nobreak \ifvmode \relax \else
      \ifdim\lastskip<1.5em \hskip-\lastskip
      \hskip1.5em plus0em minus0.5em \fi \nobreak
      \vrule height0.75em width0.5em depth0.25em\fi}

\usepackage{algpseudocode}

\begin{document}

\title{An Energy Balanced\\Dynamic Topology Control Algorithm\\for Improved Network Lifetime}
\author{\IEEEauthorblockN{Xiaoyu Chu and Harish Sethu}\\
\IEEEauthorblockA{Department of Electrical and Computer Engineering\\
Drexel University\\
Philadelphia, PA 19104-2875\\
Email: \{xiaoyu.chu, sethu\}@drexel.edu}
}

\maketitle
\thispagestyle{empty}
~\vskip 0.5in
\begin{abstract}
In wireless sensor networks, a few sensor nodes end up being
vulnerable to potentially rapid depletion of the battery reserves due
to either their central location or just the traffic patterns
generated by the application. Traditional energy management strategies, 
such as those which use topology control algorithms, reduce the energy
consumed at each node to the minimum necessary. In this paper, we use
a different approach that balances the energy consumption at each of
the nodes, thus increasing the functional lifetime of the network. We
propose a new distributed dynamic topology control algorithm called
{\em Energy Balanced Topology Control (EBTC)} which considers the
actual energy consumed for each transmission and reception to achieve
the goal of an increased functional lifetime. We analyze the
algorithm's computational and communication complexity and show that
it is equivalent or lower in complexity to other dynamic topology
control algorithms. Using an empirical model of energy
consumption, we show that the EBTC algorithm increases the lifetime of
a wireless sensor network by over 40\% compared to the best of
previously known algorithms. 
\end{abstract}

\newpage
\section{Introduction}
Wireless sensor nodes in embedded systems can sense their environment,
communicate the sensed data to each other, process the aggregated data and
cooperate on decision-making to serve a common goal. They have been deployed in a number of application areas include
traffic management, surveillance in both military and civil
settings, inventory management in businesses, and pollution monitoring
during disasters. 
However, such
tasks are possible only when the system as a whole continues to
possess enough battery resources to accomplish its objective. Any such
system typically has a few critical participating nodes that end up
expending energy faster than others, thus reducing the functional
lifetime of the system.

Extending the lifetime of a wireless sensor network (WSN) has often been
accomplished through mechanisms that improve the battery life on
each individual node (such as through reduced usage or
through improved battery technologies). But if some node will
invariably have more tasks than others (e.g., because of their
location), then, if we can evenly distribute energy consumption rate among all the sensor nodes,
the chances of system becoming non-functional will be reduced. 


In this paper, we
propose a {\em distributed topology control algorithm}
that allows each node to modulate its transmission power (thus
controlling the topology by controlling who can communicate with who
others) with the shared goal of extending the lifetime of the
system. The problem of topology control is typically constrained by
each node maximizing its lifetime while maintaining some global goal
such as connectivity. In our approach, each sensor node does not
merely try to reduce its energy consumption, but tries to balance the
energy consumption across all nodes to serve the common objective of
an increased lifetime of the system, while also preserving
connectivity.




\subsection{Related work}\label{sec:related_work}
Traditional topology control algorithms use an energy
management strategy in which each sensor node reduces its
transmission power from its maximum value to a relatively small
quantity while maintaining the connectivity of the network.
Algorithms of this class include {\em Directed
Relative Neighborhood Graph (DRNG)} \cite{LiHou2005-1313}, 
{\em Directed Local Spanning Subgraph (DLSS)} \cite{LiHou2005-1313}, \emph{ Step Topology
Control (STC)} \cite{SetGer2010} and \emph{ Cone-Based Topology Control (CBTC)}
\cite{LiHal2005}. 
The topology of the network is
determined at the very beginning of the network's life, and
remains the same throughout the network lifetime.

%

Even though these algorithms can improve the network lifetime they do
not exploit the full potential of the network because they do not
consider that, due to the geographical location of the nodes or other
application-specific reasons, some sensor nodes may have a much larger
transmission power than some other nodes or may be burdened with a
large traffic load. As a result, the energy consumption rates on these
nodes are much higher than on other nodes, resulting in an uneven
energy dissipation among the sensor nodes. This unevenness leads to a
situation where a few critical and heavily used nodes may be `dead'
before most others and the network nodes are no longer able to
cooperate meaningfully and perform the tasks needed to meet the common
objective. Meanwhile, it is possible that there are many other nodes
in the system with a large energy reserve going unused. Such a
situation requires more frequent intervention by human personnel in
monitoring and replacing nodes in the system. If the energy
consumption on the sensor nodes can be distributed more evenly, the
lifetime of the network can be extended.

The energy level on each sensor node is taken into account in \emph{Weighted
Dynamic Topology Control (WDTC)} algorithm \cite{SunYua2011}. By
assigning the edge weight based on the energy levels of the sensor
nodes on both ends of the edge, the WDTC algorithm computes a local minimum 
spanning tree based on the data exchanged. This allows the WDTC algorithm to
distribute energy consumption amongst sensor nodes more evenly and
therefore, extends the lifetime of a WSN. 

But the WDTC algorithm does not
allow for the fact that, for any sensor node, the energy consumed for
sending and receiving a packet may be different. The energy
consumed in receiving a packet is usually less than that of the
energy consumed for sending the same packet
\cite{HeiCha2000,TiaGeo2003,MilVai2004,KolPav2011}. In other words, if
the sender and the receiver node have the same amount of energy, the
sender node will very likely to be the node that runs out of energy
first. 
Therefore, the current energy level alone does not tell us enough to
estimate the remaining lifetime of a wireless sensor node.

%
%
%

Another body of dynamic topology control algorithms tackles a
specific scenario where data packets are transmitted to a specific
sink node. In the work of \cite{ShiSon2011}, a centralized dynamic
topology control algorithm is proposed which employs the Max-Flow
algorithm to help determine the network topology. Rather than adjusting the weights of edges within the
network, it adjusts the capacity of each sensor node based on the
node's current energy level and transmission power.


Other approaches to increasing the lifetime of a wireless sensor
network include grouping nodes into clusters to create a communication
hierarchy in which nodes in a cluster communicate only with their
cluster head and with only cluster heads being allowed to communicate with
other cluster heads or the sink node
\cite{AbaAna2009,KolPav2011}. A survey of topology control
algorithms can be found in \cite{MahMin2008,San2005}. 

\subsection{Contributions and Problem Statement}

In this paper, we propose a new distributed dynamic topology control algorithm
which overcomes the weaknesses discussed above of previously proposed
static protocols such as DLSS\cite{LiHou2005-1313}, STC \cite{SetGer2007} or dynamic protocols such as WDTC \cite{SunYua2011}. 
Our algorithm adapts the topology of the network based on the energy level
left on each sensor node within the network while taking into account
the actual energy consumed for sending and receiving a data
packet. By doing so, the energy consumption among the sensor nodes
is more evenly distributed, extending the lifetime of the network. We
refer to this algorithm as the {\em Energy Balanced Topology Control
  (EBTC)} algorithm.

{\em Problem statement.}
According to IEEE 802.11, 
a wireless node that receives a
frame from a sender at the medium access control (MAC) layer will
respond back directly to the sender with an acknowledgment frame at
the MAC layer. Under such a circumstance, the graph deduced from the
network has to consist of only bi-directional links. In this paper, we
abide by this standard and, accordingly, the EBTC algorithm generates
topologies that only consist of bi-directional links.

Let the graph $G(t)=(N,E(t))$ represent the topology of a wireless sensor
network at time $t$, where $N_i\in N$ represents a node within the
network with id $i$, and $(N_i,N_j)\in E(t)$ represents the fact
that node $N_j$ is within node $N_i$'s communication radius at time
$t$ and can communicate with $N_i$ directly. Since $G(t)$ only
consists of bidirectional links, then if $(N_i,N_j)\in E(t)$, we can
conclude that $(N_j,N_i)\in E(t)$. Let $S_i(t)$ denote the energy
level of node $N_i$ at time $t$. Since the lifetime of the network
is defined as the time when the first node within the network runs
out of energy \cite{SunYua2011}, we can conclude that the
lifetime of the network $T=\min(t|S_i(t)=0,N_i\in N)$. So now the
problem becomes one that tries to maximize $T$ such that the
lifetime of the WSN is maximized.

{\em Organization.} Section
\ref{sec:pseudo_code} analyzes the rationale behind the EBTC algorithm
and presents the pseudo-code. Section
\ref{sec:performance} compares the performance of
the EBTC algorithm with some of the well-cited algorithms. Our
results show that the EBTC algorithm doubles the network
lifetime compared with static topology control algorithms and is able to improve the lifetime
by more than 40\% when compared with other existing dynamic algorithms.
Section \ref{sec:conclusion}
concludes the paper.

\section{Rationale and pseudo-code }\label{sec:pseudo_code}
In this section, we will discuss the rationale behind the EBTC
algorithm and present the pseudo-code for the algorithm.

The Directed Local Spanning Subgraph (DLSS) \cite{LiHou2005-1313}
algorithm is among the well known approaches for extending the
lifetime of a WSN. By exchanging information regarding each
node's neighborhood, a sensor node is able to construct a local
minimum spanning tree employing the Local Minimum Spanning Tree
(LMST) algorithm \cite{LiHou2005-1313}. This allows each node to
determine the localized topology of the network while maintaining the connectivity of the network. Because of the simplicity
and the performance of the DLSS algorithm, in the EBTC algorithm we
employ DLSS to help determine the initial local topology of the
network.
In the DLSS algorithm, the weight of an
edge is determined by the transmission power needed in order for the
two nodes on both ends of the edge to communicate directly. In the
EBTC algorithm, we introduce a new edge weight assignment approach
which incorporates both the energy level available on the sensor nodes
and the direction of the data transmission.


\subsection{Edge weight assignment}

A glossary of terms used here is provided in Table \ref{table:term}.

Denote the minimum transmission power required for node $N_i$ to
communicate with node $N_j$ directly as $P_{i,j}$. Assume
$P_{i,j}=P_{j,i}$. Let $E_S(i,j,m)$ denote the energy required for
node $N_i$ to send a packet of size $m$ to node $N_j$, and let
$E_R(m)$ denote the energy required to receive a packet of size $m$.

If we denote the weight of edge $(N_i,N_j)$ as $w_{i,j}$, then $w_{i,j}$ can be written as:
\begin{equation}\label{equ:weight}
w_{i,j}=w_{j,i}=\max(C_{i,j}, C_{j,i})
\end{equation}
where
\begin{equation}
\begin{split}
C_{i,j}=\max\left(\frac{E_{S}(i,j,m)+E_R(m')}{S_i}, \frac{E_R(m)+E_{S}(j,i,m')}{S_j}\right)\\
C_{j,i}=\max\left(\frac{E_{S}(j,i,m)+E_R(m')}{S_j}, \frac{E_R(m)+E_{S}(i,j,m')}{S_i}\right)\nonumber
\end{split}
\end{equation}
Here, $m'$ represents
the size of the ACK message. Recall that $S_i$ denotes the energy
reserves at node $i$. 

The reasoning for this weight assignment function is as follows. The
energy consumed for successfully sending a data packet of size $m$
from node $N_i$ to node $N_j$ is the sum of the energy consumed for
sending the packet ($E_S(i,j,m)$) and for receiving the ACK for it
($E_R(m')$). The term $S_i/(E_S(i,j,m)+E_R(m'))$, therefore, indicates the
number of times that a packet of size $m$ can be successfully sent
from node $N_i$ to node $N_j$ before node $N_i$ runs out of energy.
On the other hand, since node $N_j$ has to receive the packet from
node $N_i$ and send an ACK back to node $N_i$, the energy consumed
in this process is $E_R(m)+E_S(j,i,m')$. Therefore, the term
$S_j/(E_R(m)+E_S(j,i,m'))$ indicates the number of packets that node
$N_j$ can receive from node $N_i$ before running out of energy. Both
$N_i$ and $N_j$ have to have enough energy reserves in order to
perform a successful data transmission. Therefore, in this case, the
term
$Z_{i,j}=\min(S_j/(E_R(m)+E_S(j,i,m')),S_i/(E_S(i,j,m)+E_R(m')))$
indicates the maximum number of packets that can be sent through the
link $N_i\rightarrow N_j$ before one of the nodes runs out of
energy. We define $Z_{i,j}$ as the lifetime of edge $(N_i,N_j)$.

\begin{table}[!t]
\begin{center}
\begin{tabular}{cp{6cm}}
\hline
Notation & Definition \\
\hline
$N_i$    &Sensor node with id $i$.\\
\hline
$P_i$    &Node $N_i$'s transmission power.\\
\hline
$P_{i,j}$    &Power necessary for node $N_i$ to communicate with node $N_j$ directly.\\
\hline
$S_i$    &Energy level on node $N_i$.\\
\hline
$G$    &The topology of the network when every node is transmitting at its maximum transmission power $P_{\text{Max}}$. $G=(N,E)$\\
\hline
$R_i$    &Node $N_i$'s neighbors when transmitting at $P_{\text{Max}}$. $R_i=\{N_j|(N_i,N_j)\in E\}$.\\
\hline
$V_i$    &Node $N_i$'s two-hop neighbors. $V_i=\{N_k|N_k\in R_i \text{ or } N_k\in \{R_j|N_j\in R_i\}\}$.\\
\hline
$G_i$    & Node $N_i$'s local graph containing two-hop neighbor information. $G_i=(V_i,E_i)$. The weight of the edge is calculated by Eqn.\,\,\ref{equ:weight}.\\
\hline
$E_i$    &$E_i=\{(N_j,N_k)|N_j,N_k\in V_i,(N_j,N_k)\in E\}$\\
\hline
$G_i'$        & $G_i'=(V_i,E_i')$. The subgraph of $G_i$ after DLSS algorithm.\\
\hline
$Y_i$    &Node $N_i$'s neighbors after DLSS algorithm. $Y_i=\{N_j|(N_i,N_j)\in E_i'\}$.\\
\hline
$C_{i,j}$        &The cost of link $(N_i,N_j)$\\
\hline
$Z_{i,j}$        &The lifetime of link $(N_i,N_j)$\\
\hline
$w_{i,j}$        &The weight of edge $(N_i,N_j)$. $w_{i,j}=w_{j,i}$\\
\hline
$d_{i,j}$        &Euclidean distance between node $N_i$ and $N_j$\\
\hline
$E_S(i,j,m)$        &Energy necessary to transmit a packet of size $m$ from node $N_i$ to $N_j$\\
\hline
$E_R(m)$        &Energy necessary to receive a packet of size $m$\\
\hline
\end{tabular}
\vspace{8pt}
\caption{A glossary of terms used in this section} \label{table:term}
\vspace{-25pt}
\end{center}
\end{table}
 
If all the edges' lifetimes are determined in such a fashion, then
the cost of a transmission through a particular edge can be
represented by the percentage of lifetime that may be consumed
through this transmission. In this paper, we denote this quantity as
the cost of the edge. If we let $C_{i,j}$ represent the cost of edge
$(N_i,N_j)$, then we have $C_{i,j}=1/Z_{i,j}$.

On the other hand, if the power levels of nodes $N_i$ and $N_j$ are
different and, noting the fact that the size of a data packet is
usually different from the size of its ACK, the values of $C_{i,j}$
and $C_{j,i}$ may be different. Therefore, sending a
packet along one direction may cost more than sending the same
packet in the opposite direction. Since the EBTC algorithm
aims to generate a topology which only consists of bi-directional
links, this requirement leaves the algorithm two options:
\begin{itemize}
\item Option 1: assign different weights for the edges in opposite directions.
\item Option 2: assign a unified weight for edges in opposite directions.
\end{itemize}

\begin{figure}[!t]
\begin{center}
    \subfigure[{The initial topology of the network where sensor nodes have different energy levels.}]{
       \includegraphics[width=1.55in]{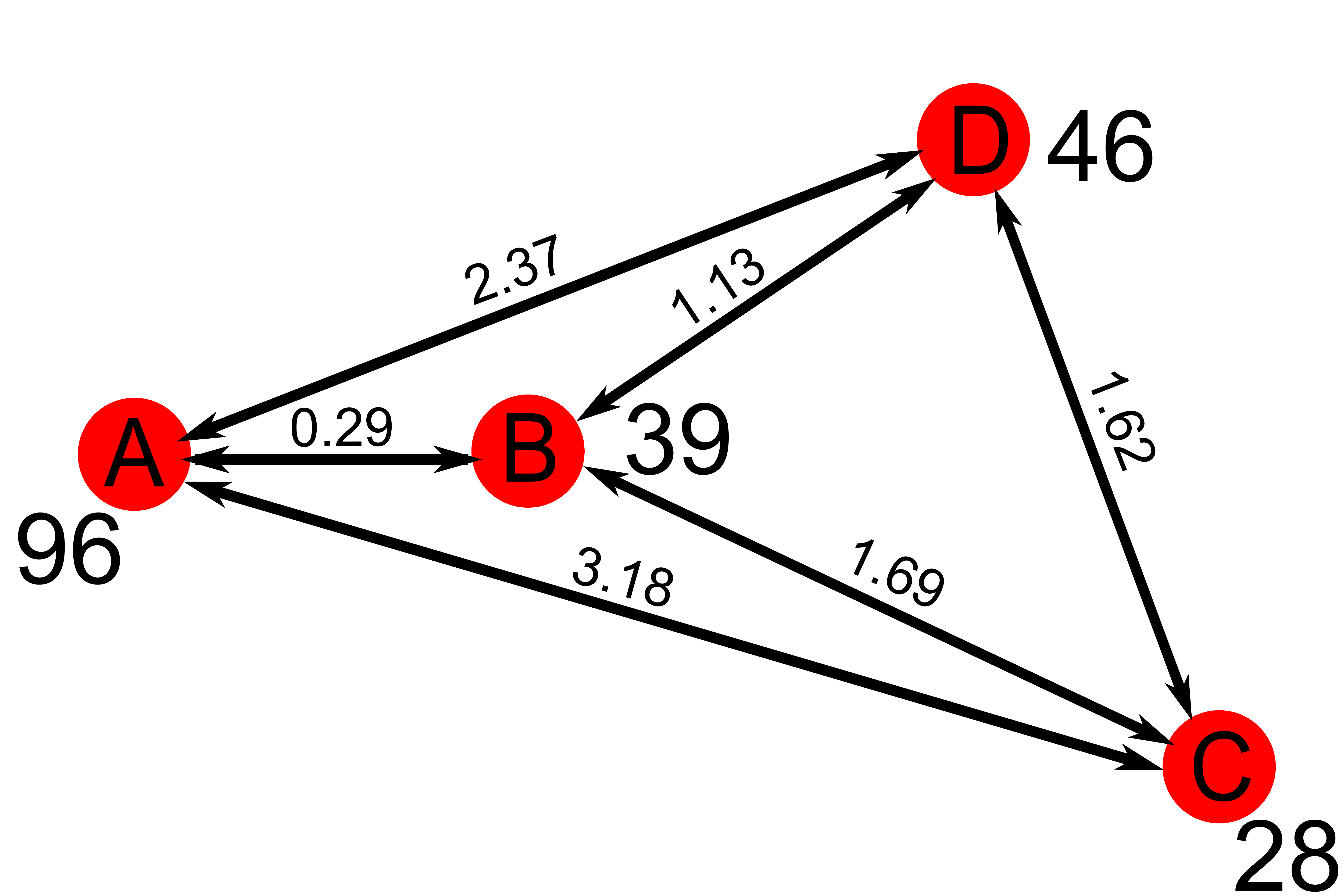}
       \label{fig:ini}
       }\hskip0.1in
    \subfigure[{The weighted graph deduced from the original topology of the network, where the edge weights are assigned according to Option 1.}]{
       \includegraphics[width=1.55in]{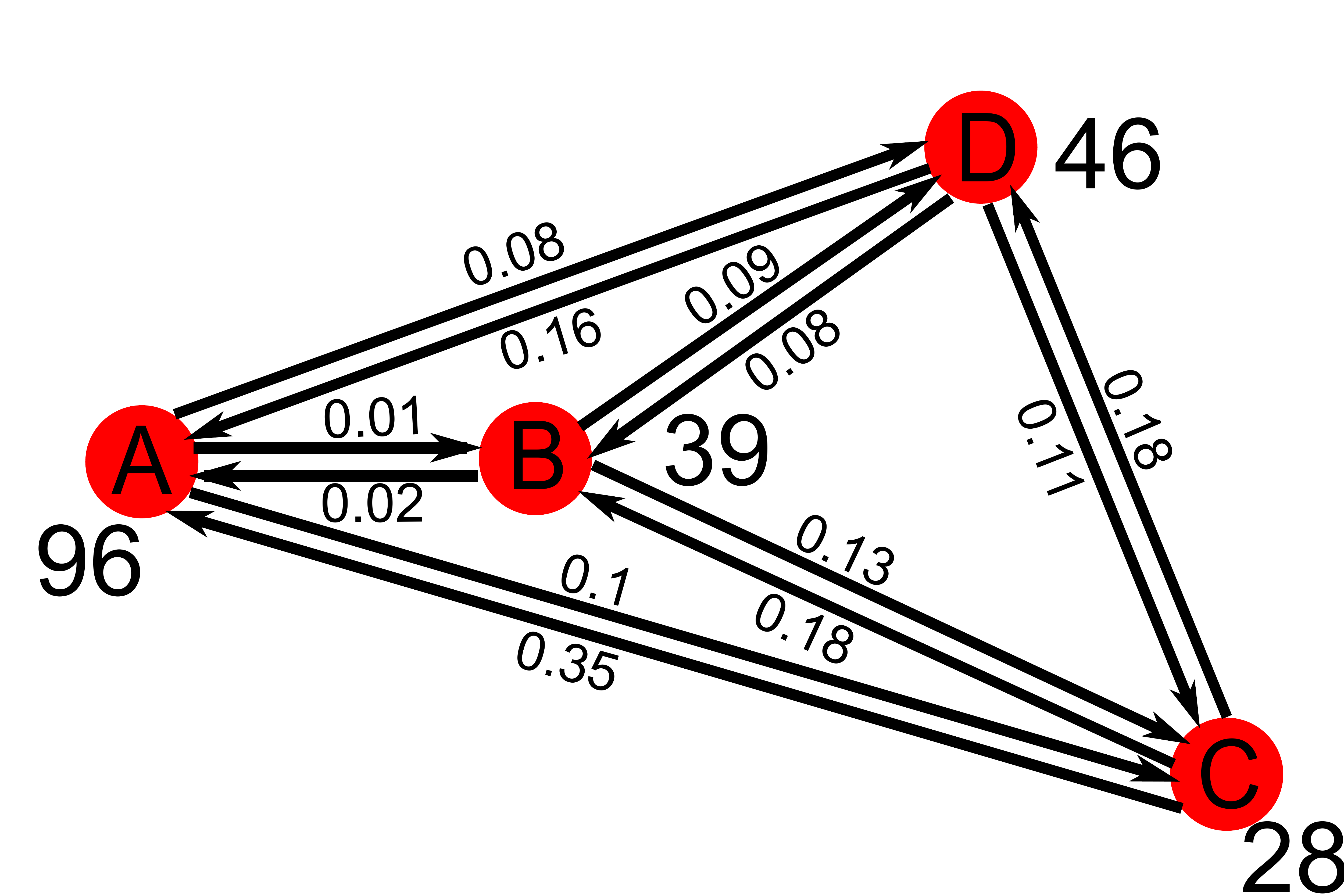}
       \label{fig:weighted}
       }\hskip0.6in
     \subfigure[{The deduced subgraph from Fig.\,\ref{fig:weighted} after the MST algorithm.}]{
       \includegraphics[width=1.55in]{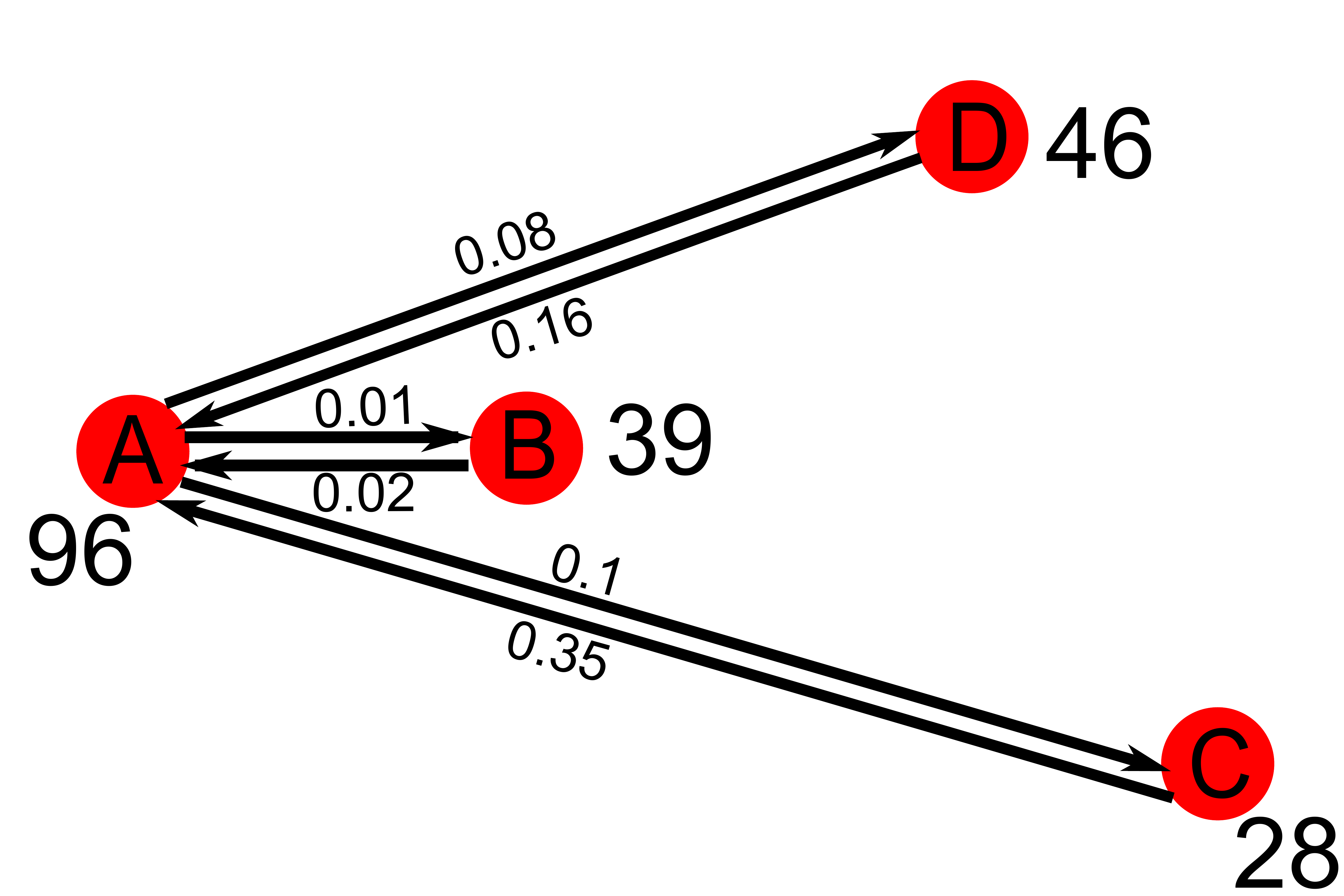}
       \label{fig:weightedChanged}
       }\hskip0.1in
     \subfigure[{The final topology of the network by employing edge weight Option 1.}]{
       \includegraphics[width=1.55in]{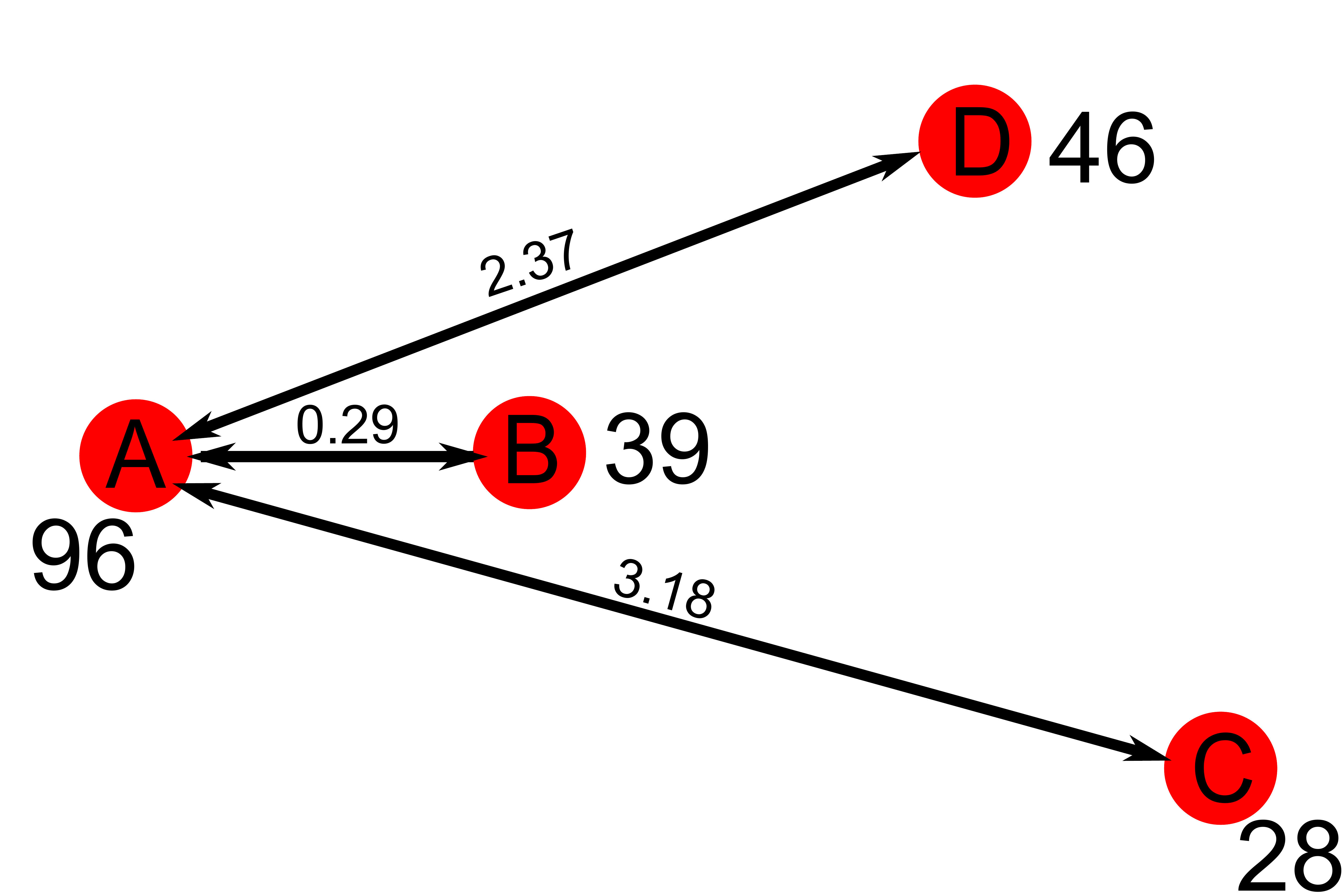}
       \label{fig:weightedFial}
       }\hskip0.1in
    \subfigure[{The weighted graph deduced from the original topology of the network, where the edge weights are assigned according to Option 2 }]{
        \includegraphics[width=1.55in]{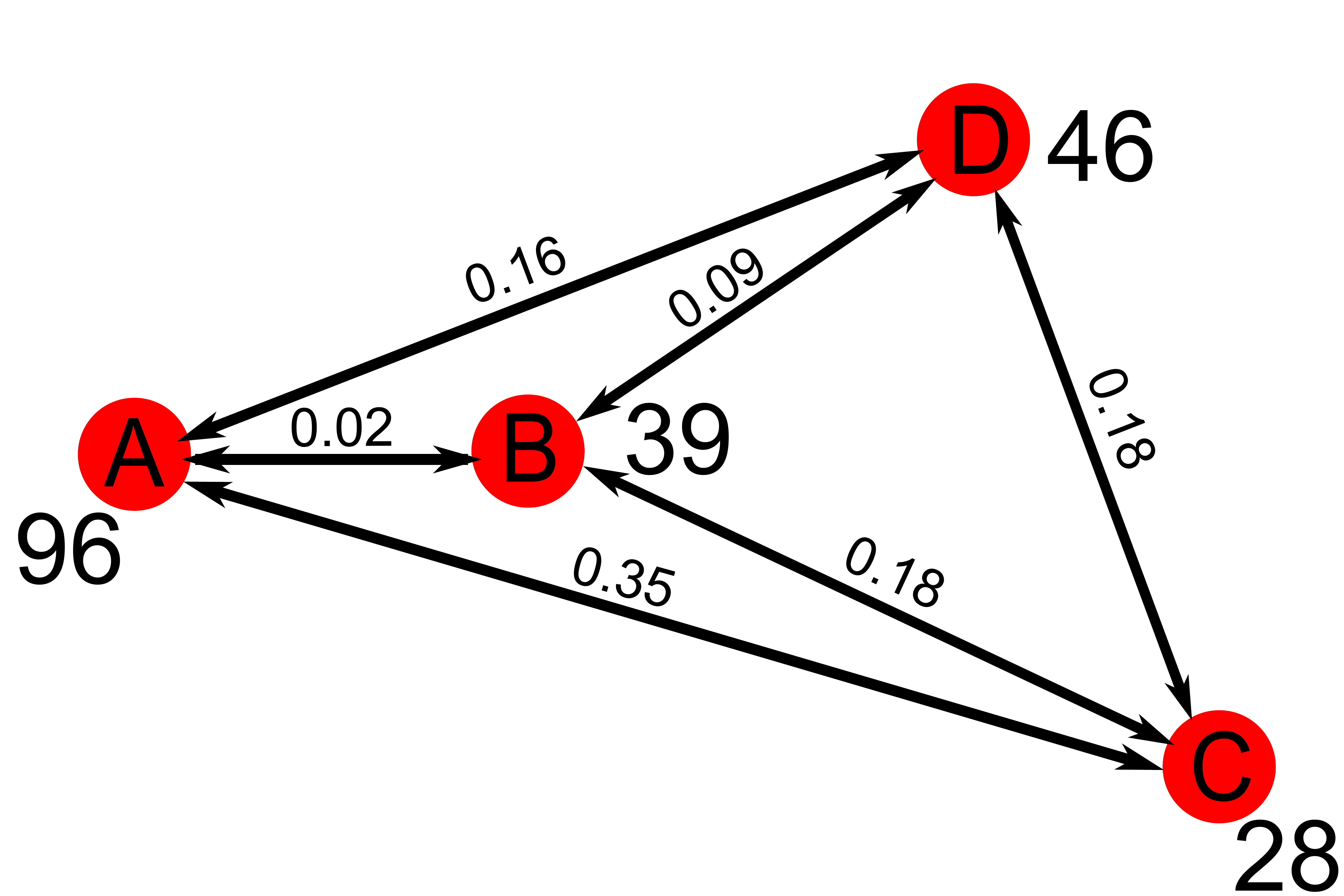}
        \label{fig:Final}
        }\hskip0.1in
    \subfigure[{The final topology of the network as the result of edge weight assignment Option 2.}]{
        \includegraphics[width=1.55in]{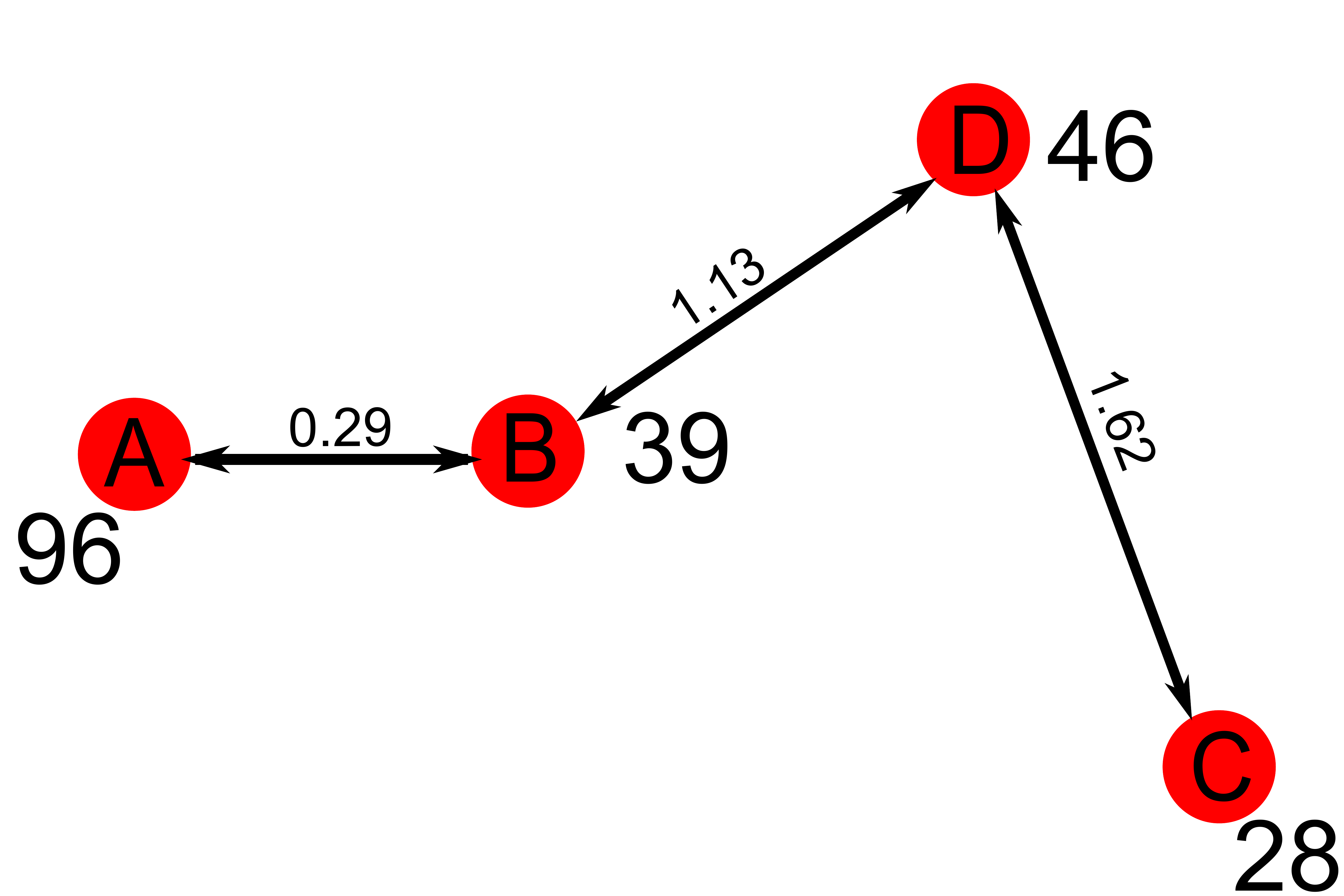}
        \label{fig:Final_after_final}
        }
   \caption{An example illustrating how the edge weights are assigned in our algorithm.}\label{fig:WeightAssignment}
\end{center}
\vspace{-20pt}
\end{figure}

In the EBTC algorithm, we adopt the second approach (Option 2). The
rationale for it is illustrated by considering the situation shown in
Fig.\,\ref{fig:WeightAssignment}.

In Fig.\,\ref{fig:WeightAssignment}, the number beside each sensor
node denotes the amount of energy available on the sensor nodes.
Suppose Fig.\,\ref{fig:ini} illustrates the topology of the network
at some time instance within the network's lifespan. The number beside each edge indicates the
minimum transmission power required for the two nodes at both ends
of the edge to be able to communicate with each other directly. It is the actual cost of communication
between any pair of nodes. Clearly, the most amount of energy is required in order for node $A$
to be able to communicate with node $C$ directly.



In Fig.\,\ref{fig:weighted}, we plot the corresponding weighted
graph based on edge weight assignment Option 1
, i.e., $w_{i,j}=C_{i,j}$. Since edges pointing at
opposite directions may have different weights according to weight
assignment Option 1, we treat each bi-directional link
within the graph as the union of two directed links of opposite
directions. The weight of each directed edge is marked beside the
specific edge. For example, transmitting through link $A\rightarrow D$
will cost at most $8\%$ of node $A$'s and node $D$'s lifetime while
transmitting the same amount of information through the reverse link
$D\rightarrow A$ will cost either node $A$ or node $D$ $16\%$ of its
remaining lifetime. Note that, $w_{i,j}$ does not necessarily equal
$w_{j,i}$. In fact, unless $S_i$ and $S_j$ are the same, $w_{i,j}$ and $w_{j,i}$ will not
be the same.

Given the topology illustrated in Fig.\,\ref{fig:weighted}, we can
run a simple Minimum-Spanning-Tree (MST) algorithm to find the
reduced topology of the network. The resulting topology of the
network is illustrated in Fig.\,\,\ref{fig:weightedChanged}.
According to the MST algorithm, first, edges $(A,B)$ and $(B,A)$ are
added to the graph containing only the nodes. Then, edges $(D,B)$ and
$(A,D)$ are included in the final topology. Note that, since the final
topology of the network has to consist of only bi-directional links,
edges $(B,D)$ and $(D,A)$ are also brought into the final topology
because of the existence of edges $(D,B)$ and $(A,D)$.

Now comes the interesting part. Note that edge $(A,C)$ has the
smallest weight of all the edges left to be tested. Since there
exists no path between node $A$ and node $C$, edge $(A,C)$ will be
selected into the final topology. Following exactly the same reasoning
that the final topology consists of only bi-directional links, edge
$(C,A)$ is brought into the graph. The resulting topology of the
network is illustrated in Fig.\,\ref{fig:weightedChanged}. At this
point, we should note that edge $(C,A)$ has the largest cost of all
the edges within the graph, and node $C$ has to transmit at its
maximum transmission power in order to stay connected with node $A$.
The final topology of the network is illustrated in
Fig.\,\ref{fig:weightedFial} where the number on each edge indicates
the actual transmission power required to communicate with the node
from the other end of the edge directly. Obviously, this is the
worst situation; the node with the least amount of energy is
transmitting at its largest transmission power and the lifetime of
the network is, therefore, greatly limited.

To avoid such a scenario, we choose edge weight assignment Option 2
such that edges in opposite directions are assigned the same
weight. The value of the edge weight equals the larger of the two
costs, i.e., $w_{i,j}=w_{j,i}=\max(C_{i,j},C_{j,i})$. In
Fig.\,\ref{fig:Final}, we plot the weighted graph based on the new
edge weight function. Again, the number on the edge indicates the
weight of the edge. For example, the weight of the edge $(A,D)$ is
chosen to be the larger value of the cost of edge $(A,D)$, which is
$0.08$, and the cost of edge $(D,A)$, which is $0.16$ as indicated in
Fig.\,\ref{fig:weighted}.

By running the MST algorithm on the new weighted graph, we are able to
determine the topology of the network as illustrated in
Fig.\,\ref{fig:Final_after_final}. The number above each edge
indicates the actual transmission power needed for the two nodes at
both ends of the link to be able to communicate with each other
directly. As indicated in Fig.\,\ref{fig:weightedFial}, node $C$ is
able to reduce its transmission power from $3.18$ to
$1.62$, node $D$ is able to halve its transmission power from $2.37$
to just $1.13$ while node $A$ is able to reduce its transmission power
from $3.18$ to just $0.29$.


\subsection{Pseudo-code}
The Energy Balanced Topology Control algorithm is described in
pesudo-code form in Algorithm \,\ref{alg:top}.

Each sensor node will execute the EBTC algorithm periodically until
it runs out of energy. At the beginning of the EBTC algorithm, each
sensor node will broadcast its current energy level at its maximum
transmission power $P_{\text{Max}}$. After collecting its neighbor
information, it will calculate each edge weight based on
Eqn.\,\,\ref{equ:weight}. The sensor node will then broadcast this
information at maximum transmission power $P_{\text{Max}}$ such
that all its neighbors will be able to collect its local information
and construct their own local graph. We refer to this process as
the {\em Collect Data} phase, which is described in lines 13--18 in
Algorithm \,\ref{alg:top}. The communication complexity in this phase
is $O(\Delta^2)$, where $\Delta$ represents the order of node $N_i$'s
degree. After receiving information from all its neighbors, node $N_i$
will construct its local graph $G_i$ which consists of two-hop
neighbor information. This process is illustrated by the function
{\em Construct Local Graph (CLG)} described in lines 19--31 in
Algorithm \,\ref{alg:top}. The order of computational complexity in
this phase is $O(\Delta^2)$.

\begin{algorithm}[!t]
\caption{Energy Balanced Topology Control (EBTC) at node
$N_i$}\label{alg:top}
\begin{algorithmic}[1]
\Require{Data packet size $m$, Maximum transmission power
$P_{\text{Max}}$}
\Ensure{$G_i'=(V_i,E_i')$, the local topology of node $N_i$}
\State Collect Data()
\State $G_i\leftarrow$CLG()
\State $G_i'=(V_i,E_i')\leftarrow$ DLSS($G_i$)
\State{$Y_i=\{N_j|(N_i,N_j)\in E_i'\}$}
\State {Broadcast $Y_i$ at $P_{\text{Max}}$}
\State {Receive $Y_j$ from $N_j\in R_i$}
\For {$N_j\in Y_i$}
\If {$N_i \not\in Y_j$}
\State{Remove $(N_i,N_j)$ from $E_i'$}
\EndIf
\EndFor
\State{$P_i=\max(P_{i,j}|(N_i,N_j)\in E_i')$}
\Statex

\Function{Collect Data}{}
\State{Broadcast id $i$ and $S_i$ at $P_{\text{Max}}$}
\State{Compile neighbor list $R_i$ }
\State{Compute $U_i=\{w_{i,j}|N_j\in R_i\}$}
\State{Broadcast $U_i$ and $R_i$ at $P_{\text{Max}}$}
\EndFunction
\Statex

\Function{Construct Local Graph (CLG)}{}
\State{$G_i=(V_i,E_i)$, $V_i\leftarrow R_i$, $E_i\leftarrow \emptyset$ }
\For{$N_j\in R_i$}
\State{$E_i\leftarrow E_i\bigcup (N_i,N_j)$, $w_{i,j}\leftarrow w_{i,j}$}
\For{$N_k\in R_j$}
\State{$E_i\leftarrow E_i\bigcup (N_j,N_k)$, $w_{j,k}\leftarrow w_{j,k}$}
\If{$N_k\not\in V_i$}
\State{$V_i\leftarrow V_i\bigcup N_k$}
\EndIf
\EndFor
\EndFor
\State\textbf{return} $G_i$
\EndFunction{}
\end{algorithmic}
\end{algorithm}

After the construction of the local graph, the DLSS algorithm is
called to reduce node $N_i$'s local topology based on the weighted
graph $G_i$. After execution of the DLSS algorithm, a subgraph is
produced and node $N_i$'s local topology is determined. At this point,
the computational complexity is the same as that of the DLSS
algorithm, which is $O(\Delta^2\log \Delta)$. Note that, at this
point, the topology of the network may not necessarily consist of only
bi-directional links. To ensure the bi-directionality of all the
edges, some additional steps have to be taken.

Given any graph, to convert it into a graph consisting of only
bi-directional links, two approaches could be taken:
\begin{itemize}
\item Remove edges that do not have a reverse link.
\item Add a reverse link to the edges that do not have one.
\end{itemize}
According to the discussion in \cite{LiHou2005-1313}, both approaches
will ensure the bi-directionality of all edges within the
network while militarizing the connectivity of the network. Since one of the goals of a topology control algorithm is to
reduce interference among transmissions, therefore, in the EBTC
algorithm, we choose to remove the edges that do not have a reverse
link. This process may also help reduce node $N_i$'s transmission
power even further.

To facilitate such an operation, node $N_i$ will broadcast
its new neighbor information $Y_i$ at power $P_{\text{Max}}$ (line 5). Upon
receiving the newly updated neighbor information from all its
neighbors in the original topology, node $N_i$ will examine the
bi-directionality of its local topology. Any edge that does not have
a reverse edge is removed from its local topology (lines 7--11) and its
transmission power is determined to be the smallest transmission
power necessary to maintain the connectivity of its local topology
(line 12).

At this point, we can conclude that the communication complexity of
the EBTC algorithm each round is the same as that of the DLSS
algorithm, which is $O(\Delta^2)$, and the computational complexity of
the EBTC algorithm is also the same as that of the DLSS algorithm,
which is $O(\Delta^2\log \Delta)$.

\section{Simulation Results}\label{sec:performance}
\begin{figure*}[!t]
\begin{center}
    \subfigure[{The lifetime for different algorithms.}]{
       \includegraphics[width=2.17in]{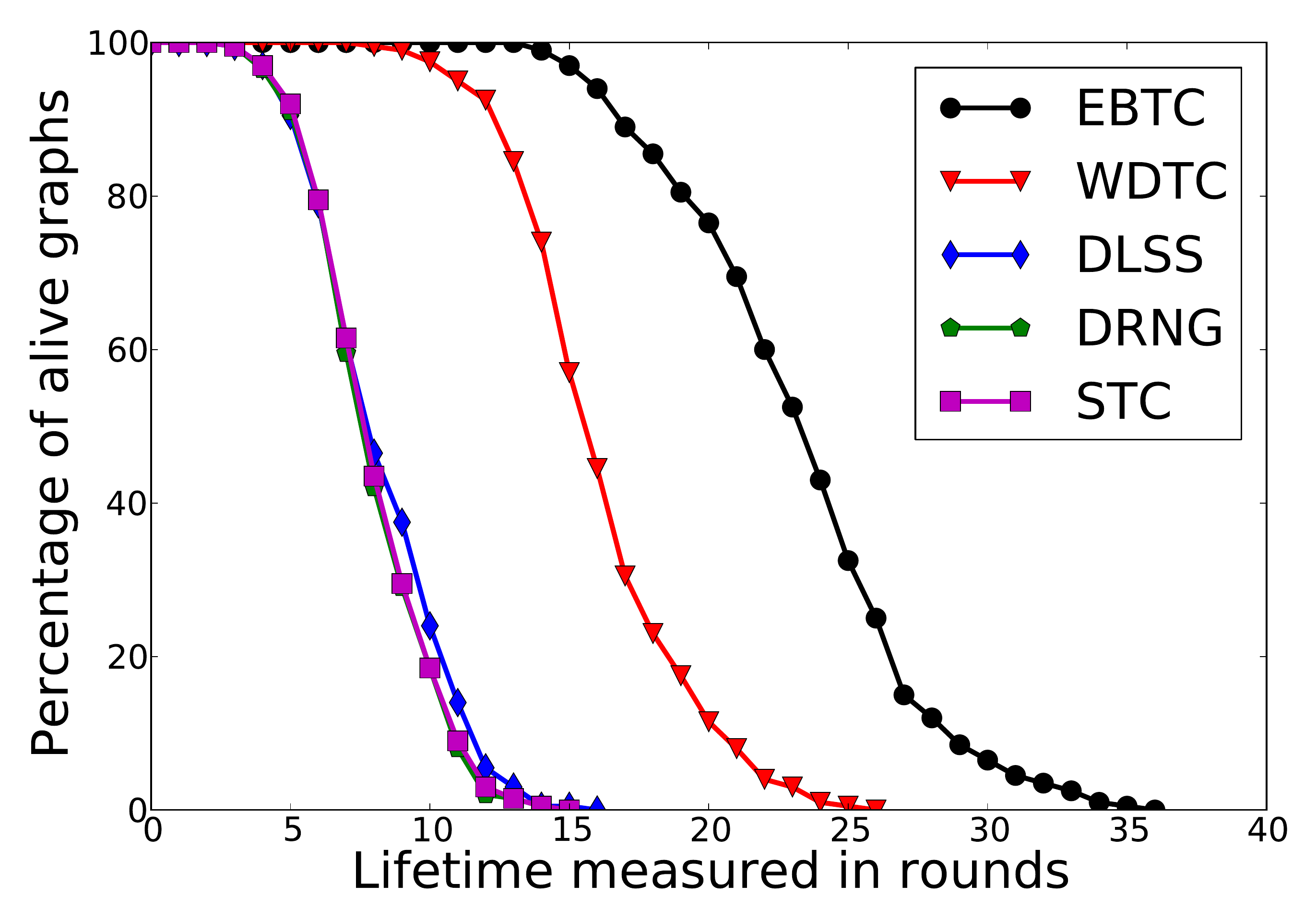}
       \label{fig:lifetime}
       }\hskip0.2in
    \subfigure[{The average transmission power per node for different algorithms plotted against time.}]{
        \includegraphics[width=2.1in]{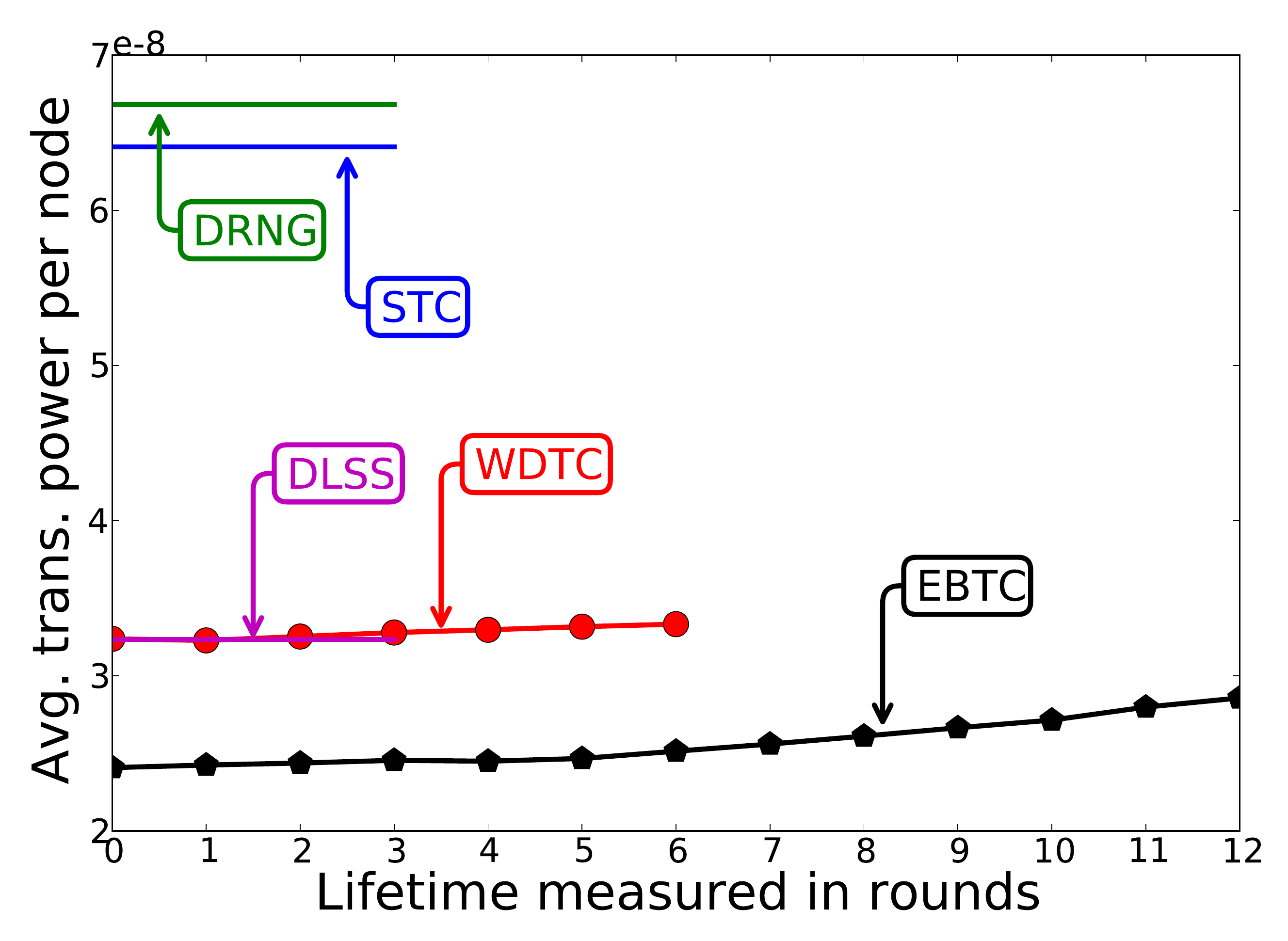}
        \label{fig:power}
        }\hskip0.2in
    \subfigure[{The average energy consumption along the minimum energy path plotted against time.}]{
      \includegraphics[width=2.17in]{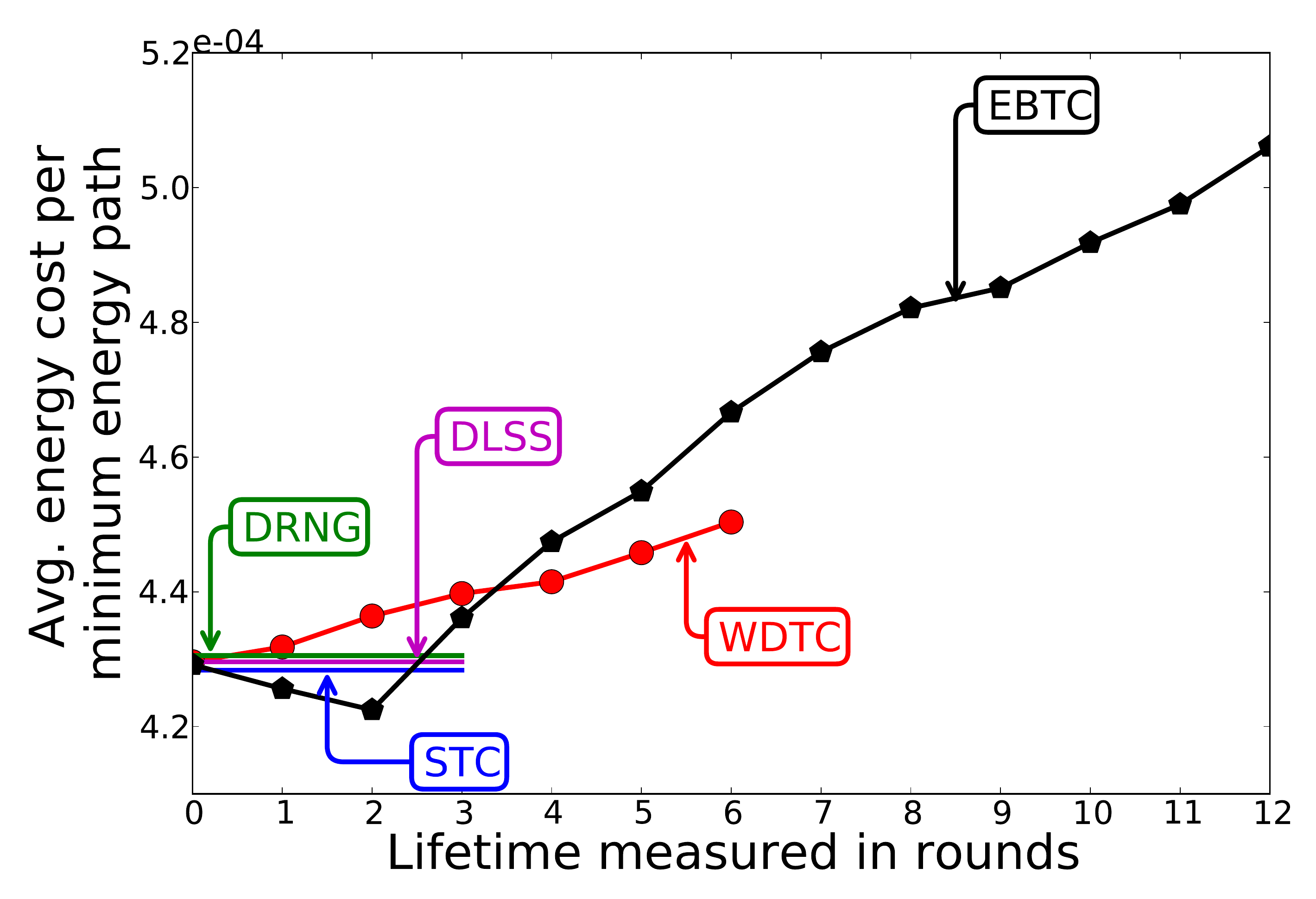}
        \label{fig:cost}
        }
   \caption{A comparative performance analysis of different algorithms
     when the routing path used is the one corresponding to the minimum
     energy consumption (i.e., packets are always routed along the
     minimum-energy path). }\label{fig:Type1}
\end{center}
\vspace{-12pt}
\end{figure*}
Our simulation is conducted in a square 1000m$\times$1000m region
within which 200 nodes are placed in random locations. Each node is
equipped with 10J of energy and has a maximum transmission power
$P_{\text{max}}$, which translates into a transmission radius of 20\%
of the width of the square region. Each data point in the results
reported here is the average of 200 randomly generated graphs. Using
the batch means method to estimate confidence intervals, we have
determined that the 95\% confidence interval is within $\pm$2\% for
each of the data points reported in our results.

In our simulation
model, we employ the TinyOS standard \cite{tinyos} for sensor node
data transmission, including its packet format for data and
acknowledgements. The energy model used in our simulation is identical to that is used
in the research literature on topology control
\cite{HeiCha2002,KolPav2011}. This model incorporates
energy consumption due to transmission, reception, and for radio
electronics in both free space and over a multi-path channel above a
certain distance threshold. Note that since the EBTC algorithm is an algorithm that takes the actual energy cost for sending and receiving a package into consideration and constructs the link weight based on this information, it is well adapted to work with any energy model selected.

In this session, we compare the performance of the EBTC algorithm
against some of the existing well-cited algorithms: DRNG \cite{LiHou2005-1313},DLSS \cite{LiHou2005-1313}, STC \cite{SetGer2010} and the
WDTC \cite{SunYua2011}. For
purposes of a meaningful comparison, all the algorithms that we have
selected for analysis in this section have a similar order of
communication and computational complexity, are fully distributed, and
only produce a topology that consists of bidirectional
links. Algorithms which cannot adapt to the IEEE 802.11 standard's
requirement that all links allow bidirectional communication, such
as those proposed in \cite{ChuSet2012}, are not included in our analysis.


The lifetime of the network is measured in `rounds'. Within each
round, each sensor node will adjust its local topology based on the
algorithm provided. Then, a designated data packet of 32 bytes will be
sent out to every other node within the network, i.e, a node will send
out $n-1$ packets each round. The lifetime of the network is therefore
indicated by the number of rounds for which network will survive
before one of its nodes runs out of energy.

In our experiment, data packets are routed through the minimum-energy
path. The simulation results are reported in
Fig.\,\,\ref{fig:Type1}. Fig.\,\,\ref{fig:lifetime} reports the
network lifetime achieved by different algorithms. For each point in
the graph, its x-axis value indicates the number of rounds that has
passed while its y-axis value indicates the percentage of graphs that
are still alive (no node has run out of energy). As can be observed
from the graph, the EBTC algorithm is able to postpone the time when
$50\%$ of the graphs tested becomes `dead' by two times when compared
with static topology control algorithms such as DLSS, DRNG and STC. When
compared with the dynamic topology control algorithm WDTC, the EBTC
algorithm is able to improve this time by about $40\%$.

Figs.\,\ref{fig:power} and \ref{fig:cost} report the changes in the
average transmission power per node and the average energy
consumption along the minimum energy path achieved by each algorithm
as time passes. For static algorithms such as DLSS, DRNG and STC,
the topology of the network is determined at the very beginning of
the network's lifetime and therefore, the network's parameters
remain the same throughout the network's lifetime. Dynamic topology
control algorithms such as WDTC and EBTC are able to adjust the
topology of the network based on the energy level available on the
sensor nodes, and therefore, produce different results.

To offer a fair comparison for static topology control algorithms, we
have only reported the performance of these algorithms until the time
when the first graph dies. This explains the limited number of rounds
for which these algorithms have data points in Figs.\,\ref{fig:power}
and \ref{fig:cost}. Nevertheless, we are able to observe that dynamic
topology control algorithms are able to offer an extended lifetime
compared with static algorithms. Since the EBTC algorithm is able to
dispense energy consumption among sensor nodes more evenly, it is able
to double the lifetime of the network which dies first among all the
networks when compared with the performance of the WDTC algorithm.

It also can be observed that as the sensor nodes gradually exhaust their
energy resources, the topology of the network changes. To extend the
lifetime of the nodes with limited resources, alternative paths are
generated by its neighboring nodes such that these nodes may be able
to reduce their transmission powers, or be able to reduce their traffic load. These changes are made usually
at the cost of some node increasing its transmission power, and
since the resulting topologies are not most cost-effective, the
average transmission power and the average cost along the
minimum-energy path gradually climb, as can be observed in
Figs.\,\ref{fig:power} and \ref{fig:cost}.

An interesting phenomenon can be observed in the performance of the
EBTC algorithm reported in Fig.\,\ref{fig:cost}. The energy cost
along the minimum energy path drops in the first two
rounds, and then gradually increases as time passes. This sudden change in
the energy cost along the minimum energy path may be explained by
the drastic change in the topology of the network. Since the EBTC
algorithm is more sensitive to the direction of the data
transmission, it is able to differentiate between the sender and the
receiver at the two ends of an edge. The EBTC algorithm, therefore, is
able to offer a better energy conservation topology compared with
other algorithms, especially when the energy levels of sensor
nodes on the two ends of an edge are different. As time passes by,
more nodes are needed to increase their transmission power so as to
compensate for the rapid energy lost on some of the sensor nodes,
resulting in the increase in both average transmission power per node
and the average energy consumption along the minimum-energy path.

\section{Conclusion}\label{sec:conclusion}
In this paper, we proposed a localized dynamic topology control
algorithm (EBTC) to help extend the lifetime of a WSN. By adapting to the energy reserves left on the sensor nodes,
the EBTC algorithm is able to distribute energy consumption more
evenly among the nodes and therefore extends the network's
lifetime. By incorporating the actual energy consumption for sending
and receiving a data packet into the decision-making process, the
EBTC algorithm is able to more accurately estimate the lifetime of a sensor node and,
therefore, greatly improves the lifetime of the network.

According to our simulation results, the EBTC algorithm is able to
increase the lifetime of the network by more than $100\%$ compared to the best
of the static algorithms and is able to extend the
lifetime of a wireless sensor network by roughly $40\%$ when compared
to other known dynamic topology control algorithms.



\bibliographystyle{IEEEtran}
\bibliography{chu-sethu}
\end{document}